\documentclass[12pt,preprint]{aastex}
\usepackage{latexsym,graphicx,natbib}

\newcommand{\beq}{\begin{equation}}
\newcommand{\eeq}{\end{equation}}

\begin{document}
\bibliographystyle{icarus}

\title{The Short Rotation Period of Nereid}

\shorttitle{Short Rotation Period of Nereid}
\shortauthors{Grav, Holman, Kavelaars}
\medskip

\author{Tommy~Grav}
\affil{\footnotesize \it Institute of Theoretical Astrophysics, University in Oslo,\\
	Postbox 1029 Blindern, 0359 Oslo, Norway (tommy.grav@astro.uio.no) \\
	 \& \\
	Harvard-Smithsonian Center for Astrophysics, \\ 
	MS51, 60 Garden Street, Cambridge MA 02138}
\email{tgrav@cfa.harvard.edu}
\author{Matthew~J.~Holman\altaffilmark{1}}
\affil{\footnotesize \it Harvard-Smithsonian Center for Astrophysics, \\
MS51, 60 Garden Street, Cambridge, MA 02138}
\email{mholman@cfa.harvard.edu}
\author{JJ~Kavelaars\altaffilmark{1}}
\affil{\footnotesize \it 
National Research Council Canada\\
5071 West Saanich Rd.\\		
Victoria, BC V9E 2E7}
\email{JJ.Kavelaars@nrc-cnrc.gc.ca}
\altaffiltext{1}{Visiting Astronomer,
 Cerro Tololo Inter-American Observatory.  CTIO is operated by the
     Association of Universities for Research in Astronomy (AURA)
Inc., under contract with the National Science Foundation.}

\medskip

\date{\rule{0mm}{0mm}}

\begin{abstract}

We determine the period, $p = 11.52 \pm 0.14$~h, and a 
light curve peak-to-peak amplitude, $a = 0.029 \pm 0.003$ magnitudes, of the Neptunian 
irregular satellite Nereid.  If the light curve variation is due 
to albedo variations across the surface, rather than solely to 
the shape of Nereid variations, the rotation period would be a factor
of two shorter.  In either case, such a rotation period and 
light curve amplitude, together with Nereid's orbital period, 
$p=360.14$~days, imply that Nereid is almost certainly in a regular 
rotation state, rather than the chaotic rotation state suggested by 
\citet{Schaefer.1988,Schaefer.2000,Dobrovolskis.1995}.

Assuming that Nereid is perfectly spherical, the albedo 
variation is $3\%$ across the observed surface. Assuming a uniform
geometric albedo, the observed cross sectional area varies by $3\%$.
We caution that the lightcurve found in this paper only sets limits 
on the combination of albedo and physical irregularity and that we cannot 
determine the orientation of Nereid's spin axis from our data.

\end{abstract}



\section{Introduction}
{\bf N~II~Nereid}, one of the irregular satellites of Neptune, was
discovered in 1949 by G.~Kuiper from McDonald Observatory
\citep{Kuiper.1949}. Nereid is physically large ($\sim 175 \pm 25$km 
radius) for an irregular moon \citep{Smith.1989,Thomas.1991}, and has
an extremely eccentric orbit ($e\sim0.75$). 

The photometric and rotational properties of Nereid are still
undetermined, despite numerous ground-based and space-based observations.  
Reported light curves give amplitudes from 
an upper limit of 0.05 magnitudes reported by \citet{Buratti.1997} to 
a 1.5 magnitude amplitude reported by \citet{Schaefer.1988}. Reported rotation 
periods range from hours to as much as a year. It should 
be noted that a recent study by \citet{Schaefer.2001} suggests that 
a large opposition effect might explain much of the controversy.
The large intra-night variations reported by \citet{Schaefer.1988} and 
\citet{Williams.1991}, however, still remain unexplained. 

The uncertainties in Nereid's rotation state would be of relatively
little concern were it not for the theories of Nereid's origin and
possible chaotic rotation state. It is suggested that Nereid formed as 
a regular satellite around Neptune but was ejected to its present orbit by
Triton after Triton was captured from heliocentric orbit and its orbit
was tidally circularized \citep{McKinnon.1984,Goldreich.1989,Banfield.1992}.  
Furthermore, it has been suggested that the reported large amplitude
photometric variations are the result of chaotic tumbling due to the
overlap of resonances between the spin and orbit periods of Nereid, similar to
that predicted \citep{Wisdom.1984} and observed 
\citep{Klavetter.1989a,Klavetter.1989b,Black.1995} for the Saturnian moon
Hyperion. \citet{Dobrovolskis.1995} has studied the effect of spin-orbit 
resonances and tidal evolution on Nereid in detail.  He suggests that tides
slowed Nereid's rotation period to a few days or weeks while Nereid was in
orbit close to Neptune.  After Nereid was scattered by Triton the
satellite has been further despun to a period of the order a month as
it reached its current 360-day orbit.  \citet{Dobrovolskis.1995} also
points out that, for rotation periods of Nereid longer than about two weeks,
Nereid is it is likely to be in spin-orbit resonance if Nereid is
nearly spherical (less than 1\%).  Otherwise, Nereid's rotation is
likely chaotic, with its period and obliquity changing from year to
year.  However, for rotation periods shorter than two weeks, Nereid is
unlikely to be in spin-orbit resonance or to be tumbling chaotically.

In this paper we report new, accurate relative photometry of
Nereid.  In the next section, we review previous observational results on the
photometry of Nereid.  In section 3,
we discuss our observations and data reduction procedures. 
In section 4, we report the characteristics of Nereid's light curve. 
In the final section, we summarize our conclusions.

\section{Previous Observations}

Kuiper's original magnitude estimate of 19.5 was
the only available photometry until \citet{Schaefer.1988} reported large 
amplitude photometric variations (1.5 magnitudes) and a possible rotational 
period of 8 to 24 hours in observations of Nereid over the period of
18-26 June of 1987. 

A number of subsequent studies found similar results.  
\citet{Bus.1988} and \citet{Bus.1989} reported photometric variations,
with a peak-to-peak amplitude of approximately $0.5$ magnitudes,
in observations covering 14 nights in June and July 1988 and in June
1989. \citet{Williams.1991} reported 1.3 magnitude amplitude
variations over 6 consecutive nights in July 1990 and argued for a $13.6$~hr
period. \citet{Schaefer.2000} reported their entire collection of 224
photometric observations of Nereid from 1987 to 1997, in which they
confirmed large brightness variations with a total amplitude of $1.83$
magnitudes on time scales ranging from hours to approximately a
year. They also reported a shift in the brightness variations, from
large amplitude rapid variations with intranight changes before $\sim
1991$ to slower, smaller amplitude variations, with no detectable
intranight changes. 

On the other hand, Voyager II, in 1989 found no brightness variations 
greater than $10\%$ \citep{Thomas.1991} and no evidence that Nereid 
is significantly aspherical, although the resolution (43.3km/pixel 
and later 61km/pixel) could not constrain this beyond the general radius
determination of $175 \pm 25$km \citep{Smith.1989}. 

\citet{Buratti.1997} observed Nereid on three nights in July 1995 with
the Palomar 5~m telescope and found no large brightness
variations, although they did report a $0.14$ magnitude decrease between
their two first nights (their first night only allowed a few images due to a 
forest fire). They adressed the discrepancy between their data set (and
that of Voyager) and the data sets reported by
\citet{Schaefer.1988}, \citet{Bus.1988} and \citet{Williams.1991},
suggesting that the large brightness variations observed were due
to significantly understated errors of the earlier observations. 

\citet{Brown.1998} observed Nereid in the R-band on two consecutive nights 
and found no variation beyond a $0.09 \pm 0.05$ magnitude increase between 
the two nights (a 3$\sigma$ result that did not include any systematic errors). 
They concluded that their data is consistent with a light curve 
with $\Delta m < 0.1$ magnitude, although a long-periodic, large amplitude 
light curve could not be ruled out.

Most recently, \citet{Schaefer.2001} used 57 V-magnitudes collected over 
52 nights in the period from June 20 to October 26 in 1998 to determine the
opposition surge of Nereid. They found a suprisingly large phase
coefficient of $0.38$ magnitudes per degree for phase angles less than 
$1^\circ$ and $0.03$ magnitudes per degree for phase angles greater than
$1^\circ$.  \citet{Schaefer.2001} noted that, although the large brightness
variations found in many of the runs (11 of 16) from 1987 to 1998 could be
explained by such an opposition surge, not all of the apparent
variation could be accounted for by phase effects alone. A closer examination
of the available data, reveal that 4 of the 5 runs that can not be explained by 
the phase effects are from 1987-1990 when Nereid was only $13-17^\circ$ away 
from the galactic center. The star densitity in these areas makes accurate 
photometry very difficult with even state of the art methods. All of these runs also 
have intranight variations, which further makes the accuracy of these observations
questionable.

\section{Observations and Data Reduction}

We observed N~II~Nereid during a pencil beam search for faint
Neptunian satellites using the 8k MOSAIC camera and a VR-filter 
\citep{Allen.2001} on the CTIO 4~m Blanco telescope on 2001 
August 9-13 and 2002 August 12-16. Nereid was only observed on one night in 2001, but in 2002
our search fields were placed such that they slightly overlapped,
ensuring that Nereid was observed on all four nights. The
exposure times used were 480 seconds with a temporal resolution of
10-15 minutes in 2001, and 20 to 40 minutes in 2002 (see Table \ref{tab:res}).  
The 2001 search strategy consisted of staying on one single field throughout 
the night, while in 2002 alternating exposures between two fields was used. 
The pointing of the CTIO 4m Blanco Telescope is accurate to about $10-20$ 
pixels, insuring that even with Nereid's motion of $\sim 15$ pixels/hour,
the moon stayed within $\sim 100$ pixels throughout the night. It is known
that the CTIO 8k MOSAIC camera causes a variation in the zeropoint across
the field-of-view (FOV). Depending on the night, Nereid moved either radially 
or tangentially acrossed the FOV. This, together with the small change in 
radial distance from the FOV center during the night, the maximum change in
zeropoint is $\sim 0.002$ magnitudes, within the statistical errors of our data.

The images were bias-subtracted, flat-fielded, and relative aperture
photometry was performed \citep{Howell.1989}. The
full-width-half-maximum (FWHM) of each image was measured ($1$ to
$1.5$ arcseconds). An aperture with radius $1.2-1.5$ times the
FWHM ($1.2$ to $2.3$ arseconds) was used to ensure the maximum signal-to-noise ratio
\citep{Dacosta.1992} and at the same time minimizing the chance of
contamination from faint background sources. The same aperture was
used on a set of 10 to 12 reference stars common to all fields
throughout a night (all the refrence stars were closer than $\sim 5$ 
arcminutes and taken from the same CCD chip that contained Nereid). 
Comparing the instrumental magnitude of Nereid to
the instrumental magnitude of the reference stars on the image and
comparing this difference with that of other images reveals any
brightening or fading throughout the night. This method does not 
require photometric conditions and efficiently removes effects due 
to airmass and transparency. To test this method we applied it to several
stars with similar brightness as Nereid in the field. The resulting 
{\it ``light curves''} were flat with a root-mean-square scatter of $0.003$
magnitudes. We take this to be our systematic error and add this to 
the formal photometric errors in quadrature. To avoid contamination 
from faint background stars and galaxies we stacked all the images from each night
and found no faint sources down to $\mbox{VR}\sim25.0$ magnitude.

The magnitude differences between the individual nights were determined
by using the procedure descibed above on one of the fields from each
night, but using 10 to 15 reference stars that were common between the two
nights compared. Thus we were able to put our nightly relative
photometry on the same relative scale for all the nights in
2002. Only a few observations of standard stars \citep{Landolt.1992} 
were performed, since the the main focus of our run a search for
Neptunian satellites.  The fact that the observations were done using a
VR-filter (centered on 6000$\mbox{\AA}$ with a width of 2000$\mbox{\AA}$) 
further complicates the situation. The standard stars used have $V-R$ 
colors between $0.49-0.54$, which is slightly higher than the color of 
Nereid at $V-R = 0.44 \pm 0.03$ \citep{Schaefer.2000}, so we used the 
$R$-magnitude given by \citet{Landolt.1992} to derive a zeropoint for 
our observations. Due to the similarity in 
colors, the wider filter lets through approximately the same relative 
amount of flux for both the standard stars and Nereid. Using the 
newly derived zeropoint on our object we get an approximate $R$-magnitude
of $\sim 18.8$.  This is consistent with the magnitudes reported by 
\citet{Schaefer.2001} after accounting for the phase effects.

Due to the size of the telescope aperture and the generally excellent
observing conditions at the Cerro Tololo site, we obtained
relative photometry of Nereid with $0.003-0.006$ accuracy (the S/N
ratio of the object was 600-700). This accuracy is significantly
better than any photometry of Nereid reported to date.

\section{Results}

\begin{figure}[p]
  \begin{center}
    \leavevmode
    \includegraphics[width=6in, height=6in]{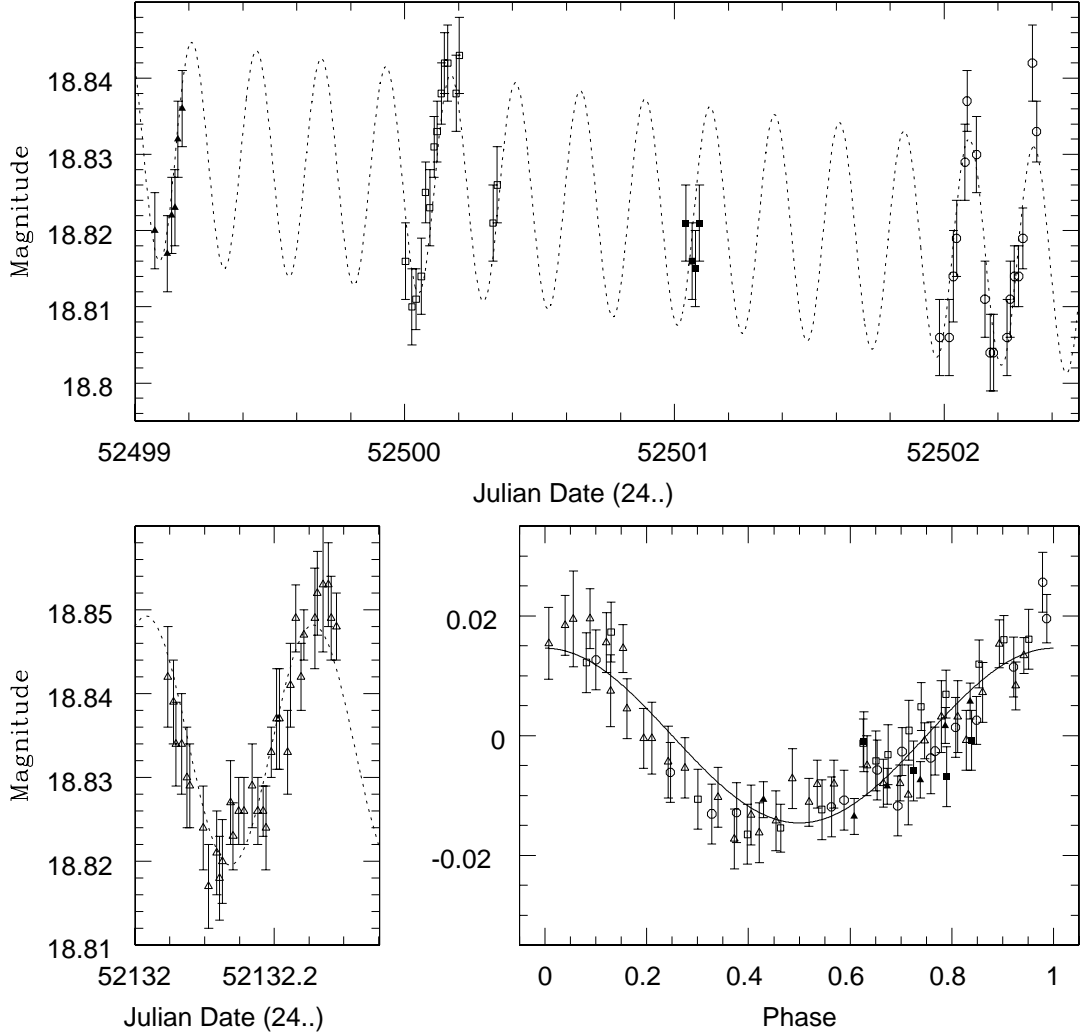}
  \end{center}
  \caption[ND]{The observations from August 2001 and 2002 are shown in
the lower left and upper panels, respectively.  The dotted
line in these two panels shows the best-fit model, which
consists of a simple sinusoid with a linear decrease in magnitude with
solar phase angle (a function of time).  The lower right panel 
shows the data from both years folded with the best-fit apparent
period.  The solid line shows the model fit.}
  \label{fig:nereid}
\end{figure}

Figure \ref{fig:nereid} shows our results, clearly indicating a periodicity 
on the order of hours. Using a Levenberg-Marquardt fitting method
\citep{Numerical.Recipes}, we fit the data with the simple model
\beq
   \Delta m = a \cos \left[ \frac{2\pi}{P} (t - t_0) \right] - k
(\alpha - \alpha_0)
\eeq  
where $t$ and $\alpha$ are the time and phase angle of the observations, and 
$a$, $P$, and $k$, are the amplitude, period, phase
coefficient, respectively. We fix the phase angle reference point,
$\alpha_0 = 0.4^\circ$. In addition to these parametes, we allow the  
sinusodial curve to move along the time axis (through letting $t_0$ be a
free parameter). We also allow the single night from 2001 to move freely along 
the magnitude axis, resulting in 6 free parameters in total. 

The fit gives a rotational period of $11.52 \pm 0.14$~hours (apparent
single harmonic period of 5.76~hours) with a peak-to-peak 
amplitude of $a = 0.029 \pm 0.003 $ magnitudes and a phase coefficient 
of $k = 0.14 \pm 0.08$ magnitudes per degree. To evaluate the fit we 
determined the chi-squared. With 68 degrees of freedom (74 observations 
minus 6 parameters) we get a chi-squared of 80.1. We further estimate the 
goodness of fit with the incomplete gamma function, $Q(0.5N,0.5\chi^2)$, 
where $N$ is the numbers of degrees of freedom \citep{Numerical.Recipes}. 
The result, $Q=0.17$, gives the probability that this variation can occur 
by chance with the given model.  We have also fit the data with higher
order harmonics to attempt to distinquish shape-induced variations
from those resulting from surface variegations, but we see no significant
improvement over the simple sinusoid with the available data.

It should be noted that the period can be well fit by values that
differ by integer multiples of $1.60\times 10^{-4}$ days or $\sim 14$
seconds, the change in period that results from one-half additional
revolution between the 2001 and 2002 observations. Obviously, we cannot 
determine the period that well with the data at hand. 
Furthermore, there is a correlation between the amplitude, $a$ and
phase coefficient, $k$.  As the period is decreased, the phase
effect coefficient increases and the amplitude decreases.  As the
period is increased, the phase coefficient decreases and the amplitude
increases. In both cases, the chi-squared increases and
$Q(0.5N,0.5\chi^2)$ decreases.  We estimate the uncertainty in the
rotation period by the limits at which $Q(0.5N,0.5\chi^2) = 0.001$ .
This yields a rotation period and phase coefficients between $11.40-11.68$ hours
and $0.19$ to $0.05$ mag/deg, respectively. Interestingly, \citet{Buratti.1997}
report a decrease of $0.05-0.025$ magitudes over a $5.5$ hour period in 
their second and third nights, although they state that this decrease was 
not statistically significant. 

The peak-to-peak amplitude of $0.029 \pm 0.003$ magnitudes does not constrain
the shape or albedo variegations of Nereid independently. Assuming that Nereid is
perfectly spherical, the albedo variation is $<3\%$ across the observed 
surface.  Recall that Voyager II constrained the brightness variations 
of Nereid over a large range of phase angles to $\le 10\%$, its radius 
to $r = 175 \pm 25$~km, and geometric albedo to  $0.180 \pm 0.005$
\citep{Smith.1989,Thomas.1991}.  Our own obervations show that,
assuming a uniform geometric albedo, the observed cross sectional area
varies by $3\%$. However, we caution that we cannot determine the orientation 
of Nereid's spin axis from our data and that if the observations are pole-on
the equatorial irregularity could well be more than $3\%$.

\section{Conclusions}

From observations on one night in August 2001 and four 
consecutive nights in August 2002 we have established the rotational period, 
$p = 11.52 \pm 0.14$ hours, and a light curve peak-to-peak amplitude, $a = 0.029 \pm 0.003$  
magnitudes, of the Neptunian irregular satellite Nereid. The peak-to-peak 
amplitude constrains the shape and/or albedo variations of Nereid. Assuming 
that Nereid is perfectly spherical, the albedo variation is $3\%$ across the 
observed surface.  Likewise, assuming a uniform geometric albedo, the 
observed cross sectional area  varies by $3\%$. Viewed from a random angle, 
this implies a nearly spherical body with a limit of $\sim 3$km out-of-round, 
based on the radius estimate from Voyager II \citep{Smith.1989,Thomas.1991}.  
Again, we caution that we cannot determine the orientation of Nereid's spin 
axis from our data.

Nereid's short rotation period and long orbital period Nereid place it near the 750:1  
spin-orbit resonance.  The phase space is essentially free of 
chaos for high rotation rates, those beyond the 40:1 spin-orbit
resonance, regardless of the shape of Nereid
\citep{Dobrovolskis.1995}.  Thus, little or no dynamical chaos is
expected in the rotation of Nereid.  Without such a chaotic region it seems
highly unlikely that Nereid could have changed it's rotational state
in recent years. Since the rotation state of Nereid is perfectly normal for a distant 
irregular satellite (cf., Jupiter VI), no implications for an unusual                          
formation history of Nereid can be drawn.

\section{Acknowledgments}

We dedicate this paper to the memory of James J. Klavetter, who
gave MH his first instruction in CCD photometry and who started this
project in 1988.  Dr. Klavetter, shortly thereafter, wisely concluded
that it would be better to wait a decade until Neptune had cleared the plane
the Milky Way than to proceed with the data in hand.

We would like to thank Drs. A. Rivkin, K. Stanek, C. Kochanek,
M. Lecar, and J. Wisdom for helpful discussions. We also thank
Dr. A. Harris for a very helpful review.  

Tommy Grav is a Smithsonian Astrophysical Observatory Pre-doctoral 
Fellow at the Harvard-Smithsonian Center for Astrophysics, Cambridge, 
USA. 

This work was supported by NASA grants NAG5-9678 and NAG5-10438.


\bibliography{ms}

\begin{thebibliography}{}

\bibitem[{Allen} {\em et~al.}(2001){Allen}, {Bernstein}, and
  {Malhotra}]{Allen.2001}
{Allen}, R.~L., G.~M. {Bernstein},\ and R.~{Malhotra} 2001.
\newblock {The Edge of the Solar System}.
\newblock {\em \apjl\/}~{\bf 549}, L241--L244.

\bibitem[{Banfield} and {Murray}(1992){Banfield} and {Murray}]{Banfield.1992}
{Banfield}, D.,\ and N.~{Murray} 1992.
\newblock {A dynamical history of the inner Neptunian satellites}.
\newblock {\em Icarus\/}~{\bf 99}, 390--401.

\bibitem[{Black} {\em et~al.}(1995){Black}, {Nicholson}, and
  {Thomas}]{Black.1995}
{Black}, G.~J., P.~D. {Nicholson},\ and P.~C. {Thomas} 1995.
\newblock {Hyperion: Rotational dynamics.}
\newblock {\em Icarus\/}~{\bf 117}, 149--161.

\bibitem[{Brown} and {Webster}(1998){Brown} and {Webster}]{Brown.1998}
{Brown}, M.~J.~I.,\ and R.~L. {Webster} 1998.
\newblock {A search for distant satellites of Neptune}.
\newblock {\em Publications of the Astronomical Society of Australia\/}~{\bf
  15}, 325--327.

\bibitem[{Buratti} {\em et~al.}(1997){Buratti}, {Goguen}, and
  {Mosher}]{Buratti.1997}
{Buratti}, B.~J., J.~D. {Goguen},\ and J.~A. {Mosher} 1997.
\newblock {No Large Brightness Variations on Nereid}.
\newblock {\em Icarus\/}~{\bf 126}, 225--228.

\bibitem[{Bus} {\em et~al.}(1988){Bus}, {Larson}, and {Singer}]{Bus.1988}
{Bus}, E.~S., S.~{Larson},\ and R.~B. {Singer} 1988.
\newblock {CCD Photometry of Nereid}.
\newblock {\em \baas\/}~{\bf 20}, 825.

\bibitem[{Bus} and {Larson}(1989){Bus} and {Larson}]{Bus.1989}
{Bus}, E.~S.,\ and S.~M. {Larson} 1989.
\newblock {CCD Photometry of Nereid}.
\newblock {\em \baas\/}~{\bf 21}, 982.

\bibitem[{Dacosta}(1992){Dacosta}]{Dacosta.1992}
{Dacosta}, G.~S. 1992.
\newblock {Basic Photometry Techniques}.
\newblock In {\em ASP Conf. Ser. 23: Astronomical CCD Observing and Reduction
  Techniques}, pp.\  90--104.

\bibitem[{Dobrovolskis}(1995){Dobrovolskis}]{Dobrovolskis.1995}
{Dobrovolskis}, A.~R. 1995.
\newblock {Chaotic rotation of Nereid?}
\newblock {\em Icarus\/}~{\bf 118}, 181--198.

\bibitem[{Goldreich} {\em et~al.}(1989){Goldreich}, {Murray}, {Longaretti}, and
  {Banfield}]{Goldreich.1989}
{Goldreich}, P., N.~{Murray}, P.~Y. {Longaretti},\ and D.~{Banfield} 1989.
\newblock {Neptune's story}.
\newblock {\em Science\/}~{\bf 245}, 500--504.

\bibitem[{Howell}(1989){Howell}]{Howell.1989}
{Howell}, S.~B. 1989.
\newblock {Two-dimensional aperture photometry - Signal-to-noise ratio of
  point-source observations and optimal data-extraction techniques}.
\newblock {\em \pasp\/}~{\bf 101}, 616--622.

\bibitem[{Klavetter}(1989a){Klavetter}]{Klavetter.1989a}
{Klavetter}, J.~J. 1989a.
\newblock {Rotation of Hyperion. I - Observations}.
\newblock {\em \aj\/}~{\bf 97}, 570--579.

\bibitem[{Klavetter}(1989b){Klavetter}]{Klavetter.1989b}
{Klavetter}, J.~J. 1989b.
\newblock {Rotation of Hyperion. II - Dynamics}.
\newblock {\em \aj\/}~{\bf 98}, 1855--1874.

\bibitem[{Kuiper}(1949){Kuiper}]{Kuiper.1949}
{Kuiper}, G.~P. 1949.
\newblock {Object near Neptune}.
\newblock {\em \iaucirc\/}~{\bf 1212}, 1.

\bibitem[{Landolt}(1992){Landolt}]{Landolt.1992}
{Landolt}, A.~U. 1992.
\newblock {Broadband UBVRI photometry of the Baldwin-Stone Southern Hemisphere
  spectrophotometric standards}.
\newblock {\em \aj\/}~{\bf 104}, 372--376.

\bibitem[{McKinnon}(1984){McKinnon}]{McKinnon.1984}
{McKinnon}, W.~B. 1984.
\newblock {On the origin of Triton and Pluto}.
\newblock {\em \nat\/}~{\bf 311}, 355--358.

\bibitem[{Press} {\em et~al.}(1995){Press}, {Teukolsky}, {Vetterling}, and
  {Flannery}]{Numerical.Recipes}
{Press}, W.~H., S.~A. {Teukolsky}, W.~T. {Vetterling},\ and B.~P. {Flannery}
  1995.
\newblock {\em {Numerical Recipes in C: The Art of Scientific Computing}}.
\newblock Cambridge University Press; Cambridge, England; 1995.~2nd ed.

\bibitem[{Schaefer} and {Schaefer}(2000){Schaefer} and
  {Schaefer}]{Schaefer.2000}
{Schaefer}, B.~E.,\ and M.~W. {Schaefer} 2000.
\newblock {Nereid Has Complex Large-Amplitude Photometric Variability}.
\newblock {\em Icarus\/}~{\bf 146}, 541--555.

\bibitem[{Schaefer} and {Tourtellotte}(2001){Schaefer} and
  {Tourtellotte}]{Schaefer.2001}
{Schaefer}, B.~E.,\ and S.~W. {Tourtellotte} 2001.
\newblock {Photometric Light Curve for Nereid in 1998: A Prominent Opposition
  Surge}.
\newblock {\em Icarus\/}~{\bf 151}, 112--117.

\bibitem[{Schaefer} and {Schaefer}(1988){Schaefer} and
  {Schaefer}]{Schaefer.1988}
{Schaefer}, M.~W.,\ and B.~E. {Schaefer} 1988.
\newblock {Large-amplitude photometric variations of Nereid}.
\newblock {\em \nat\/}~{\bf 333}, 436--438.

\bibitem[{Smith} {\em et~al.}(1989){Smith}, {Soderblom}, {Banfield}, {Barnet},
  {Beebe}, {Bazilevskii}, {Bollinger}, {Boyce}, {Briggs}, and
  {Brahic}]{Smith.1989}
{Smith}, B.~A., L.~A. {Soderblom}, D.~{Banfield}, C.~{Barnet}, R.~F. {Beebe},
  A.~T. {Bazilevskii}, K.~{Bollinger}, J.~M. {Boyce}, G.~A. {Briggs},\ and
  A.~{Brahic} 1989.
\newblock {Voyager 2 at Neptune - Imaging science results}.
\newblock {\em Science\/}~{\bf 246}, 1422--1449.

\bibitem[{Thomas} {\em et~al.}(1991){Thomas}, {Veverka}, and
  {Helfenstein}]{Thomas.1991}
{Thomas}, P., J.~{Veverka},\ and P.~{Helfenstein} 1991.
\newblock {Voyager observations of Nereid}.
\newblock {\em \jgr\/}~{\bf 96}, 19253--19259.

\bibitem[{Williams} {\em et~al.}(1991){Williams}, {Jones}, and
  {Taylor}]{Williams.1991}
{Williams}, I.~P., D.~H.~P. {Jones},\ and D.~B. {Taylor} 1991.
\newblock {The rotation period of Nereid}.
\newblock {\em \mnras\/}~{\bf 250}, 1P.

\bibitem[{Wisdom} {\em et~al.}(1984){Wisdom}, {Peale}, and
  {Mignard}]{Wisdom.1984}
{Wisdom}, J., S.~J. {Peale},\ and F.~{Mignard} 1984.
\newblock {The chaotic rotation of Hyperion}.
\newblock {\em Icarus\/}~{\bf 58}, 137--152.

\end{thebibliography}

\clearpage

\begin{table}[p]
\begin{center}
\begin{tabular}{llrrr}
 Date  & Julian & \# of Images  & Phase & Mean \\
       & Date   &  (usable) & Angle  & Anomaly \\
\hline
Aug 10 2001 & 2452132 & 33 (33) & 0.37-0.38 & $82.5^\circ$\\
Aug 12 2002 & 2452499 &  8 (~8) & 0.35-0.36 & $89.7^\circ$\\
Aug 13 2002 & 2452500 & 23 (12) & 0.38-0.40 & $90.6^\circ$\\
Aug 14 2002 & 2452501 &  4 (~4) & 0.42-0.43 & $91.5^\circ$\\
Aug 15 2002 & 2452502 & 21 (17) & 0.45-0.46 & $92.4^\circ$

\end{tabular}
   \caption[ND] {The calendar and Julian dates of the observations of 
Nereid taken with the CTIO 4-m Blanco telescope are given with the 
number of images, the solar phase angle, and the mean anomaly of
Nereid at the time of observation.
}
   \label{tab:res} 
\end{center}
\end{table}

\clearpage

\end{document}